\documentclass[twocolumn,
preprintnumbers]{revtex4}
\usepackage{graphicx}

\begin{document}
\preprint{\vtop{{\hbox{YITP-04-27}\vskip-0pt
}}}
\title{Charmed scalar mesons and related\footnote{
Talk given at the Workshop (YITP-W-03-21) on 
{\it Multi-quark hadrons; four, five and more ?}, 
Feb. 17 -- 19, 2004, Yukawa Institute for Theoretical Physics, 
Kyoto University, Kyoto}
}
\author{K. Terasaki}
\affiliation{ Yukawa Institute for Theoretical Physics, 
Kyoto University, Kyoto 606-8502, Japan
}
\thispagestyle{empty}
\begin{abstract}
Charmed scalar mesons are studied. By assigning the newly observed 
$D_{s0}^+(2.32)$ to the $I_3=0$ component, $\hat{F}_I^+$, of 
iso-triplet four-quark  mesons, $\hat{F}_I$'s, which belong to the 
scalar $[cq][\bar q\bar q], \,\,(q=u,\,d,\,s)$, multiplet, decay 
properties of the multiplet members are studied and 
existence of additional narrow scalar mesons is predicted. Decays of 
ordinary scalar $\{c\bar q\}$ mesons are also studied by comparing 
with the $K_0^*(1430)$ which has been considered as the scalar 
$\{n\bar s\},\,(n=u,\,d)$. In addition, it is demonstrated that 
contributions of four-quark mesons in hadronic weak decays of charm 
mesons, $D$ and $D_s^+$, can solve the long standing puzzle in a 
overall consistent way. 
\end{abstract}

\maketitle

As is well known, the low lying mesons including charm mesons are 
well described as the quark-antiquark systems. However, the spectrum 
of the $P$-wave states was not completed. In particular, 
in the charm sector, scalar mesons and a part of axial vector mesons 
were not observed~\cite{PDG02}. However, recently, the BABAR  
collaboration~\cite{BABAR} has observed a narrow scalar meson, 
$D_{s0}^+(2.32)$, with a mass $m_{D_{s0}}\simeq 2.32$ GeV and a width 
$\Gamma_{D_{s0}}\sim 10$ MeV (a Gaussian fit but its intrinsic width 
$\lesssim 10$ MeV) in the $D_{s}^+\pi^0$ channel. 
The CLEO~\cite{CLEO} and the BELLE~\cite{BELLE-D_s} have confirmed it 
and observed additionally axial vector 
$D_{s1}^{*+}(2.46)$~\cite{CLEO,BELLE-D_s} and 
$D_{1}^{*}(2.43)$~\cite{BELLE-D_0}, and a scalar 
$D_0^{*}(2.31)$~\cite{BELLE-D_0}. 
Thus it might be considered that all the possible $P$-wave 
$\{c\bar q\},\,(q=u,\,d,\,s)$ states were completed. 

However, since the measured value of $m_{D_{s0}}$ was much lower 
than the ones predicted by the potential model~\cite{potential} and 
the quenched lattice  QCD~\cite{quench}, various models to assign it 
have been proposed: 
(a) an iso-singlet $DK$ molecule~\cite{BCL},  
(b) an iso-singlet four-quark $\{cn\bar s\bar n\},\,\,(n=u,\,d)$, 
meson~\cite{CH}, 
(c) a mixed state of ordinary $\{c\bar s\}$ and iso-singlet 
$\{cn\bar s\bar n\}$~\cite{BPP}, 
(d) an $I_3=0$ component $\hat{F}_I^+$ of iso-triplet four-quark 
$\hat{F}_I \sim [cn][\bar s\bar n]$,  
mesons~\cite{Terasaki-D_s,Terasaki-ws},  
(e) a pole in the unitarized $DK$ amplitude~\cite{BR},  
(f) a pole in the $DK$ amplitude based on the chiral 
Lagrangian~\cite{Lutz}, 
etc., in addition to the ordinary scalar $\{c\bar s\}$~\cite{DGG} and 
the chiral partner of $D_s^+$~\cite{chiral} prior to the observation. 
In the models (a) -- (c), the narrow width of the 
$D_{s0}^+(2.32)$ is automatically satisfied since it has been 
assigned to iso-singlet states but the other molecules or four-quark 
states will be broad if they exist and have kinematically allowed 
strong decay(s). In the model (b), therefore, a narrow peak, i.e., 
the $D_{s0}^+(2.32)$, might be observed on a broad (about 100 MeV or 
more) bump arising from an $I_3=0$ component of the iso-triplet 
four-quark mesons in the $D_s^+\pi^0$ mass distribution (if the 
latter is produced sufficiently). 
The model (c) seems to be hard to distinguish from the model (a) at 
the present stage and the mixing between the $\{c\bar s\}$ and the 
$\{cn\bar s\bar n\}$ seems to be strongly dependent on the hadron 
dynamics. The model (d) will be studied later in more detail. 
The model (e) predicts that there exist a scalar $DK$ bound state and 
a broad resonance dominated by the ordinary scalar $\{c\bar s\}$ and 
that there exist two broad scalar resonances in the $D\pi$ channel. 
The model (f) which will be studied in the next talk expects 
existence of a charmed ($C=1$) scalar meson with an exotic 
combination of iso-spin and strangeness quantum numbers, 
$(I,S)=(0,-1)$, in addition to a normal $(I,S)=(0,+1)$ and two 
$(I,S)=({1\over 2},0)$ scalar mesons.     

After the observations of the $D_{s0}^+(2.32)$, the mass of the 
scalar $\{c\bar s\}$ system has been calculated from various 
approaches such as a modified potential model~\cite{mod-potential}, 
lattice QCD simulations~\cite{Bali,lattice}, a bag model~\cite{bag}, 
a quark-meson model~\cite{quark-meson}, a QCD sum rule~\cite{QCDSR}, 
a light-cone oscillator model~\cite{lightcone}, a HQET sum 
rule~\cite{HQETSR}, etc. However, almost all the models still have 
provided mass values of the scalar $\{c\bar s\}$ higher than the 
measured $m_{D_{s0}}\simeq 2.32$ GeV. 

The production rate of $D_{s0}^+(2.32)$ in $B$ decays also has been 
studied by using the factorization prescription~\cite{factorization}
and it has been concluded~\cite{Wang} that it would be nearly equal 
to that of $D_{s}^+$ and larger by about a factor ten than the 
measured ones if it is the scalar $\{c\bar s\}$ state while the rate 
for a molecule or a four-quark state would be consistent with 
experiments. 

Before studying four-quark mesons, we review very briefly potentials, 
\begin{equation}
V_{qq}({\bf r})=\sum\Lambda_i\Lambda_i v({\bf r}), \,\,\,
V_{q\bar q}({\bf r})=-\sum\Lambda_i\Lambda_i v({\bf r}), 
\end{equation}
between two quarks and between a quark and an antiquark, respectively, 
mediated by a vector meson with an extra $SU(3)$ degree of 
freedom~\cite{Hori} corresponding to the color. Although they have 
been studied much earlier than the discovery of the 
color~\cite{color}, the results which are summarized in Table~1 
are instructive. 
\begin{center}
\begin{quote}
{Table~1. 
Potentials mediated by a vector meson with $SU(3)$ "color". 
}
\end{quote}\vspace{3mm}

\begin{tabular}
{c | c | c | c | c}
\hline
&\multicolumn{2}{|c|}
{$qq$}
&\multicolumn{2}{c}{$q\bar q$}
\\
\hline
 & & & & \vspace{-4mm}\\
$SU_c(3)$
&
$\bf{\bar 3}$
&
$\bf{6}$
&
$\bf{8}$
&
$\bf{1}$
\\
 & & & & \vspace{-5mm}\\
\hline
 & & & & \vspace{-4mm}\\
\vspace{0mm}
Potential
&$\displaystyle{-{8\over 3}\langle{v}\rangle}$
&\,\,\,$\displaystyle{4\over 3}\langle{v}\rangle$
& \,\,\,$\displaystyle{{2\over 3}\langle{v}\rangle}$ 
& $\displaystyle{-{16\over 3}\langle{v}\rangle}$
\vspace{-4mm}
\\
 & & & & \vspace{-0mm}\\
\hline
\end{tabular}
\end{center}
\begin{center}
\begin{quote}
{Table~2. Ideally mixed scalar $[cq][\bar q\bar q]$ mesons, where $S$ 
and $I$ denote strangeness and $I$-spin. The number with ($\dagger$) 
is used as the input data. 
}
\end{quote}
\vspace{0.3cm}

\begin{tabular}
{c | c | c | c | c }
\hline
 & & & & \vspace{-4mm}\\
$\,\, S\,\,$
&$\,\, I=1\,\,$
&$\,\,\displaystyle{I={1\over 2}}\,\,$
&$\,\, I=0\,\,$
&$\,\,$Mass(GeV)$\,\,$
\\
 & & & & \vspace{-4mm}\\
\hline
 & & & & \vspace{-4mm}\\
$1$
&
$\hat F_I$ 
&
&
$\hat F_0$
&
{\hspace{3mm}2.32($\dagger$)}
\\
\hline
 & & & & \vspace{-4mm}\\
0
&
&
\begin{tabular}{c}
{$\hat D$}\\
{$\hat D^s$}
\end{tabular}
&
&\begin{tabular}{c}
{2.22}\\
{2.42}
\end{tabular}
\\
\hline 
 & & & & \vspace{-4mm}\\
-1
&
&
&
$\hat E^0$
&{2.32}
\\
\hline 
\end{tabular}
\end{center}

Now four-quark $\{qq\bar q\bar q\}$ mesons can be classified into the 
following four types~\cite{Jaffe}, 
\begin{equation}
\hspace{-0mm}\{qq\bar q\bar q\}=
\underbrace{[qq][\bar q\bar q]
\oplus(qq)(\bar q\bar q)}_{J^P=0^+,\,1^+,\,2^+}\oplus
\underbrace{
\{[qq](\bar q\bar q)\pm (qq)[\bar q\bar q]\}, 
}_{J^P=1^+}                   
\label{eq:four-quark}
\end{equation}
where the square brackets and the parentheses imply that 
the wave functions are anti-symmetric and symmetric, respectively, 
under exchanges of the flavors between them. The first two can have 
the spin-parity $J^P=0^+,\,1^+,\,2^+$ but the last two have only 
$J^P=1^+$. However, we are now interested in scalar mesons 
so that we concentrate on the first two. There are two ways to 
obtain color singlet four-quark states, i.e., to take the color 
$SU_c(3)$ ${\bf \bar 3}\times {\bf 3}$ and 
${\bf 6}\times {\bf \bar 6}$. As seen before, the force between two 
quarks (or between two antiquarks) is attractive when they are of 
${\bf \bar 3}$ (or ${\bf 3}$) but repulsive when they are of {\bf 6} 
(or ${\bf \bar 6}$) while the force between a quark and an antiquark 
is attractive and much stronger when they are of color singlet. 
Therefore, it is expected that the scalar $[qq][\bar q\bar q]$ mesons 
of ${\bf \bar 3}\times {\bf 3}$  of $SU_c(3)$ can be the lowest lying 
four-quark  mesons. (However, these two, i.e., 
${\bf \bar 3}\times {\bf 3}$ and ${\bf 6}\times {\bf \bar 6}$ of the 
color $SU_c(3)$, can mix with each other.) In fact, the MIT bag model 
with the bag potential and a spin-spin force shows that 
the $[qq][\bar q\bar q]$ mesons which make a ${\bf 9}$-plet of 
the flavor $SU_f(3)$ are the lowest lying states~\cite{Jaffe}. 
Such a multiplet may be realized by the observed $a_0(980)$, 
$f_0(980)$, $\sigma(600)$~\cite{PDG02} and $\kappa(800)$~\cite{E791} 
as suggested in Ref.~\cite{Jaffe}. 
To distinguish four-quark mesons with the same flavor and spin but 
different color configurations 
(dominantly ${\bf \bar 3}\times{\bf 3}$ of lower mass and 
dominantly ${\bf 6}\times{\bf \bar 6}$ of higher mass), 
we put $\ast$ on the symbols of the heavier class of four-quark 
mesons in accordance with Ref.~\cite{Jaffe}. 

Replacing one of $q$'s in the $[qq][\bar q\bar q]$ by the $c$ quark, 
we can obtain scalar $[cq][\bar q\bar q]$ mesons. We list the lighter 
class [dominantly ${\bf \bar 3}\times{\bf 3}$ of $SU_c(3)$] 
of ideally mixed $[cq][\bar q\bar q]$ mesons in Table~2, where the 
mass values have been estimated by assigning the newly observed 
$D_{s0}^+(2.32)$ to the $\hat{F}_I^+$, using the quark counting 
with $\Delta_s=m_s - m_n \simeq 0.1$ GeV and 
$m_{\hat{F_I}}=m_{D_{s0}}\simeq 2.32$ GeV as the input data. 
The $\hat{F}_I$ and $\hat {F}_0$ are the $I=1$ and $I=0$ components,  
respectively. The $\hat D$ and $\hat D^s$ are two different 
iso-doublets, where the latter contains an $(s\bar s)$ pair. 
The $\hat E^0$ is the exotic scalar meson with $C=1$ and $S=-1$, 
i.e., $\hat{E}^0\sim [cs][\bar u\bar d]$. 

We now study decay rates of the above $[cq][\bar q\bar q]$ mesons. 
The rate for the decay, 
$A({\bf p}) \rightarrow B({\bf p'}) + \pi({\bf q})$, 
is given by 
\begin{eqnarray}
&&\Gamma(A \rightarrow B + \pi)
=\Biggl({1\over 2J_A + 1}\Biggr)  
\Biggl({q_c\over 8\pi m_A^2}\Biggr) 
\nonumber\\
&&\hspace{25mm}
 \times
\sum_{spins}|M(A \rightarrow B + \pi)|^2 , 
\label{eq:rate}
\end{eqnarray}
where $J_A$, $q_c$ and $M(A \rightarrow B + \pi)$ denote the spin 
of the parent $A$, the center-of-mass momentum of the final $B$ and 
$\pi$ mesons and the decay amplitude, respectively. To calculate 
the amplitude, we use the PCAC (partially conserved axial-vector 
current) hypothesis and a hard pion approximation in the infinite 
momentum frame (IMF), i.e., ${\bf p}\rightarrow\infty$~\cite{suppl}. 
In this approximation, the amplitude is evaluated at a little
unphysical point, i.e., $m_\pi^2 \rightarrow 0$. By assuming that 
the $q^2$ dependence of the amplitude is mild as was in the old
current algebra~\cite{MP}, it is given by
\begin{equation}
M(A \rightarrow B + \pi) 
\simeq \Biggl({m_A^2 - m_B^2\over f_\pi}\Biggr)
\langle{B|A_{\bar \pi}|A}\rangle ,                         
\label{eq:amp}
\end{equation}
where $A_\pi$ is the axial counterpart of the isospin, $I(=V_\pi)$. 
The {\it asymptotic matrix element} of $A_\pi$ (matrix elements of 
$A_\pi$ taken between single hadron states with infinite momentum), 
$\langle{B|A_\pi|A}\rangle$, gives the dimensionless $AB\pi$ coupling 
strength. 

We parameterize later the asymptotic matrix elements of $A_\pi$ and 
$A_K$ using the asymptotic flavor symmetry, which is, roughly 
speaking, flavor symmetry of asymptotic matrix elements. (Asymptotic 
flavor symmetry and its fruitful results were reviewed in 
Ref.~\cite{suppl}.) However, 
the asymptotic flavor symmetry may be broken. The measure of the 
(asymptotic) flavor symmetry breaking is given by the form factor, 
$f_+(0)$'s, of related vector currents at the zero momentum 
transfer squared ($q^2=0$). The estimated values of $f_+(0)$'s are 
\begin{eqnarray}
f_+^{(\pi K)}(0)&&= 0.961 \pm 0.008, \label{eq:LR}\\
f_+^{(\bar K D)}(0)&&= 0.74 \pm 0.03,  \label{eq:PDG96}\\
{f_+^{(\pi D)}(0)\over f_+^{(\bar K D)}(0)}
&&= 1.00 \pm 0.11 \pm 0.02, \label{eq:E687} \\
&&= 0.99\pm 0.08,  \label{eq:CLEO97}
\end{eqnarray}
where the above values of the form factors, 
Eqs.(\ref{eq:LR}) -- (\ref{eq:CLEO97}), have been given in 
Refs.~\cite{LR} --~\cite{CLEO97}, respectively. They suggest that 
the asymptotic flavor $SU_f(3)$ symmetry works well while the 
asymptotic $SU_f(4)$ is broken to the extent of 20 -- 30 $\%$. 
In fact, the asymptotic $SU_f(4)$ symmetry has predicted the 
rates~\cite{suppl,HOS}, 
$\Gamma(D^{*+}\rightarrow D^0\pi^+) \simeq 96$ keV and 
$\Gamma(D^{*+}\rightarrow D^+\pi^0) \simeq 42$ keV, 
which are larger by about $40\,\,\%$ than 
$\Gamma(D^{*+}\rightarrow D^0\pi^+)= 65 \pm 18$ keV and 
$\Gamma(D^{*+}\rightarrow D^+\pi^0)= 30 \pm 8$ keV 
from the measured decay width~\cite{CLEO-D^*}, 
$\Gamma_{D^{*\pm}}=96 \pm 4 \pm 22$ keV, 
and the branching fractions compiled in Ref.~\cite{PDG02}. The above 
suggests that the size of the asymptotic matrix elements 
\begin{center}
\begin{quote}
{Table~3. Dominant decays of scalar $[cq][\bar q\bar q]$
mesons and their estimated widths. The measured width, 
$\Gamma(\hat F_I^+\rightarrow D_s^+\pi^0) \sim 10$ MeV, 
is used as the input data. 
The decays into the final states between angular brackets are not 
allowed kinematically as long as the parent mass values in the 
parentheses are taken.
}
\end{quote}
\vspace{0.2cm}

\begin{tabular}
{c c c }
\hline
& & \vspace{-4mm}\\
\begin{tabular}{c}
Parent \\
(Mass in GeV)
\end{tabular}
&$\quad$Final State$\quad$
&\begin{tabular}{c}
Width \\
(MeV)
\end{tabular}
\\
\hline
& & \vspace{-4mm}\\
\begin{tabular}{c}
$\hat F_I^{++}(2.32)$ \\
$\hat F_I^+(2.32)$\\
$\hat F_I^0(2.32)$
\end{tabular}
&
\begin{tabular}{c}
$D_s^+\pi^+$\\
$D_s^+\pi^0$\\
$D_s^+\pi^-$
\end{tabular}
& 10
\\
& & \vspace{-4.5mm}\\
\hline
& & \vspace{-4mm}\\
{$\hat D^+(2.22)$}
&\begin{tabular}{c}
{$D^0\pi^+$}\\
{$D^+\pi^0$}
\end{tabular}
&
\begin{tabular}{c}
{10}\\
{5}
\end{tabular}
\\
{$\hat D^0(2.22)$}
&\begin{tabular}{c}
{$D^+\pi^-$}\\
{$D^0\pi^0$}
\end{tabular}
&
\begin{tabular}{c}
10\\
5
\end{tabular}
\\
\hline 
& & \vspace{-4mm}\\
{$\hat D^{s+}(2.42)$}
&$D^+\eta$ 
&-- 
\\
& & \vspace{-4mm}\\
{$\hat D^{s0}(2.42)$}
&$D^0\eta$ 
&-- 
\\
\hline 
& & \vspace{-4mm}\\
{$\hat F_0^+(2.32)$}
&\begin{tabular}{c}
{$<D_s^+\eta>$}\\
$D_s^+\pi^0$
\end{tabular}
&\begin{tabular}{c}
--\\
($I$-spin viol.)
\end{tabular}
\\
& & \vspace{-4.5mm}\\
\hline 
& & \vspace{-4mm}\\
{$\hat E^0(2.32)$}
&
$<D\bar K>$
&
--
\\
& & \vspace{-4.5mm}\\
\hline 
\end{tabular}\vspace{2mm}\\
\end{center}
of axial 
charge $A_\pi$ between charmed meson states will be smaller by about 
$20\,\,\%$ than the ones in the asymptotic symmetry limit. 

Asymptotic matrix elements including four-quark meson states have 
been parameterized previously~\cite{charm88,charm93,charm99} by using 
asymptotic flavor $SU_f(3)$ symmetry. We here list the related ones, 
\begin{eqnarray}
&&\langle{D_s^+|A_{\pi^-}|\hat F_I^{++}}\rangle 
= \sqrt{2}\langle{D_s^+|A_{\pi^0}|\hat F_I^{+}}\rangle 
= \langle{D_s^+|A_{\pi^+}|\hat F_I^{0}}\rangle
 \nonumber\\
&&=-\langle{D^0|A_{\pi^-}|\hat D^{+}}\rangle            
= 2\langle{D^+|A_{\pi^0}|\hat D^{+}}\rangle 
            \nonumber\\
&&=-2\langle{D^0|A_{\pi^0}|\hat D^{0}}\rangle 
= -\langle{D^+|A_{\pi^+}|\hat D^{0}}\rangle. 
                                                 \label{eq:axial-ch}
\end{eqnarray} 
Inserting Eq.(\ref{eq:amp}) with Eq.(\ref{eq:axial-ch}) into 
Eq.(\ref{eq:rate}), we can calculate approximate rates for the 
allowed two-body decays. Here we equate the calculated rate for 
the $\hat F_I^+\rightarrow D_s^+\pi^0$ decay to the measured width 
of the $D_{s0}^+(2.32)$, i.e., 
$\Gamma(\hat F_I^+\rightarrow D^+_s\pi^0) \sim 10\,\,{\rm MeV}$,  
since we do not find any other decays which can have sizeable rates, 
and then use it as the input data when we estimate the rates for the 
other decays. The results are listed in Table~3. However, the 
numerical values should not be taken too literally since the 
intrinsic width of the $D_{s0}^+(2.32)$ as the input data is still 
not definite. (We expect that it will be in the region between a few 
and $\sim 10$ MeV. Such a narrow width is understood by the small
overlapping of wavefunctions between the initial and final 
states~\cite{Terasaki-ws}.) The calculated widths of $\hat F_I$ and 
$\hat D$ are in the region, $\sim$ (10 -- 15) MeV, so that they will 
be observed as narrow resonances in the $D_s^+\pi$ and $D\pi$
channels, respectively. The mass of $\hat D^s$ is approximately on 
the threshold of the iso-spin conserving $\hat D^s\rightarrow D\eta$, 
so that it is not clear if they are kinematically allowed. Besides, 
the decays are sensitive to the $\eta$-$\eta'$ mixing scheme which 
is still model dependent~\cite{Feldmann}.  Therefore, we need more 
precise and reliable values of $m_{\hat D^s}$, $\eta$-$\eta'$ 
mixing parameters and decay constants of $\eta$ to obtain a definite 
result. The $[cq][\bar q\bar q]$ multiplet contains the exotic state 
$\hat E^0$ with $C=-S=1$ whose mass is expected to satisfy 
approximately $m_{\hat{F}_I}\simeq m_{\hat{F}_0}\simeq m_{\hat{E}^0}$ 
from the simple quark counting as in Table~2. If it is the case, 
the $\hat{E}^0$ cannot decay through strong interactions or through 
electromagnetic interactions but only through weak 
interactions~\cite{exotic}. 

We now study decay widths of the ordinary scalar $\{c\bar s\}$ and 
$\{c\bar n\}$ mesons comparing with the $K_0^{*}(1.43)$ which has 
been considered as the $^3P_0\,\,\{n\bar s\}$ state~\cite{CT}. 
Substituting the measured values~\cite{PDG02}, 
$\Gamma(K_0^{*}\rightarrow all)=294\pm 23$ MeV 
and 
${\rm Br}(K_0^{*}\rightarrow K\pi)=93 \pm 10\,\,\%$,  
into Eq.(\ref{eq:rate}) and using Eq.(\ref{eq:amp}), we obtain 
$|\langle{K^+|A_{\pi^+}|K_0^{*0}}\rangle| \simeq 0.29$, 
where we have used the iso-spin $SU_I(2)$ symmetry which is always 
assumed in this talk. In the asymptotic $SU_f(4)$ symmetry 
limit~\cite{suppl,Hallock}, we obtain 
\begin{eqnarray}
&&
\langle{D^+|A_{\pi^+}|D_0^{*0}}\rangle 
=2\langle{D^+|A_{\pi^0}|D_0^{*+}}\rangle \nonumber\\
&&
=-2\langle{D^0|A_{\pi^0}|D_0^{*0}}\rangle 
=\langle{D^0|A_{K^-}|D_{s0}^{*+}}\rangle\nonumber\\
&&
=\langle{D^+|A_{\bar K^0}|D_{s0}^{*+}}\rangle 
=\langle{K^+|A_{\pi^+}|K_0^{*0}}\rangle,  
\end{eqnarray}
where $D_0^{*}\sim \{c\bar n\}$ and $D_{s0}^{*+}\sim \{c\bar s\}$. 
(When we take account of the about 20 $\%$ breaking of the 
asymptotic $SU_f(4)$ symmetry,) the sizes of the above asymptotic
matrix elements are estimated as 
\begin{eqnarray}
&&|\langle{D^+|A_{\pi^+}|D_0^{*0}}\rangle| 
=|2\langle{D^+|A_{\pi^0}|D_0^{*+}}\rangle| \nonumber\\
&&
=|-2\langle{D^0|A_{\pi^0}|D_0^{*0}}\rangle|
=|\langle{D^0|A_{K^-}|D_{s0}^{*0}}\rangle| \nonumber\\
&&=|\langle{D^+|A_{\bar K^0}|D_{s0}^{*+}}\rangle|  \nonumber\\
&&
=|\langle{K^+|A_{\pi^+}|K_0^{*0}}\rangle|(\times 0.8)
\simeq 0.23.  
\end{eqnarray}
It is expected that a sum of the rates for the 
$D_0^{*0}\rightarrow D^+\pi^-$ and $D^0\pi^0$ decays saturates 
approximately the total decay rate of $D_0^{*0}$. The iso-spin 
symmetry leads to 
$\Gamma(D_0^{*0}\rightarrow (D\pi)^0)
=\Gamma(D_0^{*+}\rightarrow (D\pi)^+)$.  
The decays, $D_{s0}^{*+}(2.45)\rightarrow (DK)^+$'s, also saturate 
approximately the total decay rate of $D_{s0}^{*+}$. If we take
tentatively $m_{D_0^{*}}\simeq 2.35$ GeV and 
$m_{D_{s0}^{*}}\simeq m_{D_0^{*}} + \Delta_s\simeq 2.45$ GeV which 
are around the average values predicted by the potential 
model~\cite{potential} and the quenched lattice QCD~\cite{quench}, 
we can obtain 
\begin{eqnarray}
&&
\Gamma_{D_0^{*}(2.35)} \simeq 90\,\,[\times (0.8)^2]\,\,{\rm MeV},   
\nonumber\\
&&
\Gamma_{D_{s0}^{*+}(2.45)}
\simeq   70\,\,[\times (0.8)^2]\,\,{\rm MeV},
\end{eqnarray}
where we have replaced $\pi$ by $K$ in Eqs.(\ref{eq:rate}) and 
(\ref{eq:amp}) when we obtain the second equation. 

Although the BELLE collaboration~\cite{BELLE-D_0} has recently 
reported that a charmed scalar resonance $D_0^0(2.31)$ with a mass 
$2308\pm 60$ MeV and a width $279 \pm 99$ MeV has been observed in 
the $D^+\pi^-$ channel and claimed that the result is consistent 
with the conventional $D_0^{*0}\sim\{c\bar u\}$ state, some comments 
on the above result and claim are now in order~\cite{charm-scalar}. 
In Ref.~\cite{BELLE-D_0}, it has been tried to fit four different 
model amplitudes to the measured $D^+\pi^-$ mass distribution and 
to search for the most likely solution. In all the amplitudes, 
however, only one scalar meson pole has been taken into account so 
that the $\chi^2$ value has been not sufficiently small even in the 
most likely solution which provided the above mass and width. In 
particular, significant deviations between the most likely solution 
and the measured mass distribution are seen in the broad scalar meson 
region. Therefore, it is expected that a much better fit to the 
measured $D\pi$ mass distribution will be obtained if an extra scalar 
meson pole is additionally taken into account in the model amplitude. 

The fitted mass value $m_{D_0}\simeq 2.31$ GeV of the scalar meson 
$D_0^0(2.31)$ is a little lower than the one of $D_0^*$ from a 
quenched relativistic lattice QCD~\cite{quench} and an unquenched 
but static one~\cite{Bali}, 
$m_{D_0^*}({\rm lattice})\sim 2.33$ GeV,  
while it is much lower than the one from the potential 
models~\cite{potential,mod-potential}, 
$m_{D_0^*}({\rm potential})\sim 2.4$ GeV.  
On the other hand, it is is too high when it is compared with the 
previously observed $D_{s0}^+(2.32)$. Namely, if it is assumed that 
these two are the $^3P_0\,\,\{c\bar s\}$ and $\{c\bar n\}$, it is not 
natural that they are approximately degenerate but the mass 
difference between them should be nearly equal to 
$\Delta_s \simeq 100$ MeV. (It is hard to expect the above 
degeneracy, $m_{D_{s0}} -  m_{D_{0}} \ll \Delta_s$, unless they have 
partners, i.e., extra scalar meson(s), to mix with and unless the 
dynamics of the mixings are very much different from each other.  
However, in the case of Ref.~\cite{BELLE-D_0}, the model amplitudes 
include only one scalar meson pole but no extra scalar meson as the 
partner to mix with.)  So, it is natural to consider that these two 
have different structure, for example, one is the ordinary 
$\{c\bar q\}$ and the other is a four-quark or a molecule. 

The width of $D_0(2.31)$ given by the BELLE collaboration 
was anomalously broad since, as seen before, the width of $D_0^*$ 
has been expected to be $\Gamma_{D_0^*}\sim (60 - 90)$ MeV by 
comparing with the $K_0^*(1.43)$ which has been considered as the 
$^3P_0\,\,\{n\bar s\}$ state. 
Therefore our scenario~\cite{charm-scalar} is that there coexist two 
scalar states in the  region of the broad bump around 2.31 GeV 
observed by the BELLE~\cite{BELLE-D_0} and that the heavier one is 
the ordinary $^3P_0\,\,\{c\bar n\}$ with a mass $\sim 2.35$ GeV and 
a width $\sim (60 - 90)$ MeV and the other is the four-quark 
$\hat D\sim [cn][\bar n\bar n]$ with $m_{\hat D}\simeq 2.22$ 
GeV and $\Gamma_{\hat D}\sim 15$ MeV~\cite{Terasaki-D_s}. 
It is desired that the BELLE collaboration will reanalyze the 
measured $D\pi$ mass distribution by using a model amplitude with 
two scalar meson poles. The strange counterpart $D_{s0}^{*+}$ of the 
$D_0^{*}$ will be around 2.45 GeV and its dominant decays are  
$D_{s0}^{*+}\rightarrow D^0K^+$ and $D^+K^0$ 
so that its width is $\sim$ (40 -- 70) MeV as seen before. It will be
observed as an ordinary resonance around 2.45 GeV in $(DK)^+$ 
mass distributions with high statistics. 

So far we have studied classification of the new resonances and 
strong decays of the members of the multiplet including them. 
For the production rate of the new resonance, $D_{s0}^+(2.32)$, in 
$B$-decays, we have referred to a review paper~\cite{Wang} which 
has concluded that the factorization provides too big rates if the 
$D_{s0}^+(2.32)$ is the scalar $\{c\bar s\}$ while, if it is 
a four-quark or a molecule, its production rate is consistent with 
experiments, i.e., the production rate for the four-quark mesons in 
$B$-decays will be much smaller than that of the $\{c\bar s\}$ 
mesons. Therefore, we expect that the existence of four-quark mesons 
will be confirmed in experiments with higher statistics in future, 
although we have no evidence for a peak in the 
$D_s^+\pi^\mp$~\cite{CLEO-exotic,CDF}, $D^0\pi^\mp$ and $D^+\pi^\mp$  
mass distributions~\cite{CDF} at the present stage. 

Now we study a possible role of scalar four-quark 
$[qq][\bar q\bar q]$ and $(qq)(\bar q\bar q)$ mesons in hadronic weak 
decays of charm mesons. For our basic idea on hadronic weak 
interactions, see Ref.~\cite{NOVA}, although hadronic weak decays of 
$B$ mesons have been studied in the paper. For our technical details 
to study hadronic weak decays of charm mesons, see 
Ref.~\cite{charm99} and references quoted therein. We start with 
the assumption that the decay amplitude can be given by a sum of 
factorizable and non-factorizable ones. The factorizable amplitude 
is calculated in the so-called BSW scheme~\cite{BSW} in which, 
to apply the factorization, the BSW Hamiltonian, $H_w^{\rm BSW}$, 
has to be prepared by applying the Fierz reshuffling to the 
conventional effective weak Hamiltonian, $H_w$, 
\begin{equation}
H_w \rightarrow H_w^{\rm BSW} + \tilde H_w,  \label{eq:Fierz}
\end{equation}
where 
\begin{eqnarray}
&&H_w \simeq {G_F\over \sqrt{2}}
\Bigl\{c_1Q_1^{(s'c)} + c_2Q_2^{(s'c)} +\cdots\Bigr\} + h.c. ,
                                             \nonumber \\
&&H_w^{\rm BSW} \simeq {G_F\over \sqrt{2}}
\Bigl\{a_1Q_1^{(s'c)} + a_2Q_2^{(s'c)} +\cdots\Bigr\} + h.c.,
\nonumber\\
&&\hspace{10mm}
a_1 = c_1 + {c_2 \over N_c} \gg a_2 
= c_2 + {c_1\over N_c},                                             
\nonumber\\
&&\hspace{-0mm}
Q_1^{(s'c)} = :(\bar ud')_L(\bar s'c)_L:,\quad  
Q_2^{(s'c)} = :(\bar s'd')_L(\bar uc)_L: \,\,    
\nonumber
\end{eqnarray}
with $(\bar qq)_L = \bar q\gamma_\mu(1 - \gamma_5)q$. 
Here $c_1$ and $c_2$ are the Wilson coefficients with hard gluon
corrections, and $N_c$ the color degree of freedom. As seen in 
Eq.~(\ref{eq:Fierz}), we inevitably have an 
extra term, $\tilde H_w$, which is given by a color singlet sum of 
colored current products, i.e., 
\begin{eqnarray}
&&\tilde H_w \simeq {G_F\over \sqrt{2}}
\Bigl\{c_2\tilde Q_1^{(s'c)} 
       + c_1\tilde Q_2^{(s'c)} +\cdots\Bigr\} + h.c.,  \nonumber\\
&&\hspace{10mm}
\tilde Q_1^{(s'c)} 
             = 2\sum_a:(\bar ut^ad')_L(\bar s't^ac)_L :,   
\nonumber\\
&&\hspace{10mm}
\tilde Q_2^{(s'c)} 
             = 2\sum_a:(\bar s't^ad')_L (\bar ut^ac)_L :, \nonumber
\end{eqnarray}
when we obtain the $H_w^{\rm BSW}$, where $t^a$'s are the generators
of the color $SU_c(3)$. Although it has been taken away
in the BSW scheme, we consider that it provides the non-factorizable  
amplitude which will be controlled by dynamics of hadrons and that 
the non-factorizable amplitude can play an important role in hadronic 
weak interactions of $K$, charm mesons and some of $B$ decays in 
which some selection rules such as the color (and/or 
\begin{center}
\begin{quote}
{Table~4. Branching ratios ($\%$) for the 
$D\rightarrow PP,\,(P=\pi,\,K)$ decays. 
(1) is given by the factorized amplitudes only, (2) by a sum of 
factorized and non-factorizable amplitudes, where the latter contains 
the continuum contribution and the poles of the glue-rich scalar 
and the scalar hybrid meson, and (3) by the four-quark meson pole 
amplitudes in addition to the ones in (2).}
\end{quote}
\vspace{3mm}

\begin{tabular}
{l  c  c  c  l}
\hline\vspace{-4.5mm}\\
$\hspace{5mm}\displaystyle{\rm Decays}\,\,$ 
&$\hspace{3mm}\displaystyle{(1)}\,\,$  
&$\hspace{3mm}\displaystyle{(2)}\,\,$ 
&$\hspace{3mm}\displaystyle{(3)}\,\,$  
& $\hspace{5mm}\displaystyle{{\cal B}_{\rm exp}}\hspace{0mm}$
\\
\hline
\vspace{-4mm}\\
$ \hspace{-0mm}\displaystyle{D^+\rightarrow \bar K^0\pi^+}$ 
& $\displaystyle{3.27}$  & 1.06 & $\displaystyle{2.72}$ 
& $\displaystyle{2.71\hspace{1.7mm} \pm 0.20}$ 
\\ 
\hline
\vspace{-4.5mm}\\
$\hspace{-0mm}\displaystyle{D^0\rightarrow K^-\pi^+}$ 
& $\displaystyle{2.41}$ & 9.19 & $\displaystyle{3.83}$ 
& $\displaystyle{3.83\hspace{1.7mm} \pm 0.09}$
\\
\hline
\vspace{-4.5mm}\\
$\hspace{-0mm}\displaystyle{D^0\rightarrow \bar K^0\pi^0}$ 
&$\displaystyle{0.00}$ & 3.71 & $\displaystyle{2.31}$ 
& $\displaystyle{2.30\hspace{1.7mm}\pm 0.22}$
\\
\hline
\vspace{-4.5mm}\\
\vspace{0.2mm}$\displaystyle{D_s^+\rightarrow\bar K^0K^+}$  &
$\displaystyle{0.20}$ & 7.28 & $\displaystyle{3.50}$
& $\displaystyle{3.6\hspace{3.6mm}\pm 1.1}$
\\ 
\hline
\vspace{-4.5mm}\\
$\hspace{-0mm} \displaystyle{D^0\rightarrow \pi^-\pi^+}$ 
& $\displaystyle{0.15}$ 
& 0.34
& $\displaystyle{0.14}$
& $\displaystyle{0.143\pm 0.007}$
 \\ 
\hline
\vspace{-4.5mm}\\
$\hspace{-0mm} \displaystyle{D^0\rightarrow \pi^0\pi^0}$ 
& $\displaystyle{0.00}$ 
& 0.11 
& $\displaystyle{0.09}$
& $\displaystyle{0.084 \pm 0.022}$
\\ 
\hline
\vspace{-4.5mm}\\
$\hspace{-0mm}\displaystyle{D^+\rightarrow \pi^0\pi^+}$ 
& $\displaystyle{0.12}$ & 0.12 & $\displaystyle{0.23}$ 
& $\displaystyle{0.25\hspace{1.7mm} \pm 0.07}$
\\ 
\hline
\vspace{-4.5mm}\\
$\hspace{-0mm}\displaystyle{D^0\rightarrow K^-K^+}$ 
& $\displaystyle{0.19}$  
& 0.68
& $\displaystyle{0.42}$
& $\displaystyle{0.412 \pm 0.014}$
 \\ 
\hline
\vspace{-4.5mm}\\
$\displaystyle{D^0\rightarrow \bar K^0K^0}$ 
& $\displaystyle{0.00}$  
& 0.04
& $\displaystyle{0.04}$
& $\displaystyle{0.071 \pm 0.019}$
 \\ 
\hline
\vspace{-4.5mm}\\
$\hspace{-0mm}\displaystyle{D^+\rightarrow \bar K^0K^+}$ 
& $\displaystyle{0.47}$ 
& 1.00
& $\displaystyle{0.52}$
&$\displaystyle{0.57 \hspace{1.7mm}\pm 0.06}$
\\ 
\hline
\vspace{-4.5mm}\\
\vspace{0.2mm}$\displaystyle{D_s^+\rightarrow \pi^+K^0}$ 
& $\displaystyle{0.17}$ 
& 0.17
& $\displaystyle{0.05}$
&$\displaystyle{< \,0.8}$
\\ 
\hline
\vspace{-4.5mm}\\
$\hspace{-0mm}\displaystyle{D_s^+\rightarrow \pi^0K^+}$ 
& $\displaystyle{0.00}$ 
& 0.07
& $\displaystyle{0.06}$
&\hspace{8mm} ---
 \\ \hline
\end{tabular}
\vspace{3mm}\\
\end{center}
helicity) 
suppressions work. We will estimate contributions of the 
non-factorizable amplitudes using a hard pion 
technique~\cite{suppl,hard-pion} in the infinite momentum frame 
which is an innovation of the old current algebra~\cite{MP}. In this 
approximation, the non-factorizable amplitude is given by a sum of 
all possible pole amplitudes and the so-called equal-time commutator 
(ETC) term which arises from the continuum contribution mediated by 
multi-hadron intermediate states~\cite{MP}. Among the possible pole 
amplitudes, contributions of the ordinary excited meson states 
are neglected since wave function overlappings between the excited 
states and the external states in two body decays of charm mesons 
under consideration will be small and, in particular, the value of 
wavefunction of orbitally excited state at the origin, 
$\Psi(0)_{L\neq 0}$, is expected to be small. In the $u$-channel, 
all contributions of the excited states are neglected since their 
contributions are expected to be small, while, in the $s$-channel, 
a part of (the heavier class of) scalar four-quark mesons can 
contribute to the $s$-channel pole amplitudes of the spectator decays 
(if they exist) since the related four-quark mesons are expected to 
have their masses close to the parent charm masses, $m_D$ and 
$m_{D_s}$. For example, the mass of $\hat\sigma^{s*}$ which can 
contribute to the $D^0\rightarrow K^+K^-$ is expected to be very 
close to $m_D$ and therefore it can play an important role in 
the $D^0\rightarrow K^+K^-$ decay. However, it cannot contribute to 
the $D^0\rightarrow \pi^+\pi^-$ in which the corresponding pole is 
given by $\hat\sigma^*$ but $m_{\hat\sigma^*}\ll m_D$, so that, in 
the latter, the $\hat\sigma^*$ contribution would be less important. 
In this way, we may be able to find a solution to the long standing 
puzzle~\cite{PDG02}, 
\begin{equation}
{\Gamma(D^0\rightarrow K^+K^-) \over 
\Gamma(D^0\rightarrow \pi^+\pi^-)} = 2.88 \pm 0.15, 
                                                    \label{eq:puzzle}
\end{equation}
in a overall consistent way~\cite{charm99}. In the annihilation 
decays (in the weak boson mass $m_W\rightarrow\infty$ limit), scalar 
hybrid $\{q\bar qg\}$ mesons can contribute to their $s$-channel pole 
amplitudes. The penguin term can induce scalar glue-ball (or 
glue-rich scalar meson) contributions. The factorized and 
non-factorizable amplitudes, in which contributions of hybrid mesons 
were not taken into account, have been given explicitly in 
Ref.~\cite{charm99}. 

Although the amplitudes obtained in this way includes many unknown 
parameters, we evaluated numerically and compared with experiments 
assuming that the form factors, $f_+(0)$'s, of charm changing 
vector currents satisfy the $SU_f(3)$ symmetry as seen before and 
their $q^2$ dependence is given by a monopole form as usual, taking 
the measured value of $f_+^{(\bar KD)}(0)$, using (and changing a 
little) the mass values of four-quark mesons predicted in 
Ref.~\cite{Jaffe} and treating other parameters as adjustable ones, 
and then reproduced fairly well the measured branching ratios for 
two body decays of charm mesons~\cite{charm99}. We now improve the 
above result taking additionally account of contributions of scalar 
hybrid mesons which have not been considered in our previous studies. 
Before doing this, we list the parameters involved: 
the form factor, $f_+^{(\bar KD)}(0)$; the coefficients, $a_1$ and 
$a_2$, in the $H_w^{\rm BSW}$ at the scale $\mu\sim m_c$ which might 
be some what different from the ones given by the perturbative QCD 
because of so-called final state interactions controlled by 
dynamics of hadrons~\cite{NOVA}; the asymptotic 
matrix element, 
$\langle{\pi^+|\tilde H_w|D_s^+}\rangle$;  
the parameters, $k_a^*$, $k_s^*$, $k_H$ 
and $f_g$, describing pole 
contributions (given by products of asymptotic matrix elements of 
$A_\pi$ and $\tilde H_w$, 
$\langle{\pi^+|A_\pi|n}\rangle\langle{n|\tilde H_w|D_s^+}\rangle$, 
in the unit of $\langle{\pi^+|\tilde H_w|D_s^+}\rangle$) 
of $n=[qq][\bar q\bar q]$, $(qq)(\bar q\bar q)$, $\{q\bar qg\}$ 
(scalar hybrid mesons) and $S^*$ (the glue-rich scalar meson), 
respectively;  
the relative phase $\delta$ between the factorized and the 
non-factorizable amplitudes; the phases arising from non-resonant 
meson-meson interactions, 
$\delta (\pi\pi)_0$, $\delta (K\bar K)_0$, $\delta (K\bar K)_1$, 
$\delta (\pi K)_{1/2}$ and $\delta (\pi K)_{3/2}$; 
the masses and widths of heavier class of scalar four-quark 
$[qq][\bar q\bar q]$ and $(qq)(\bar q\bar q)$, and hybrid 
mesons. For the nearest glue-rich scalar meson, $S^*$, which we will 
assign to $f_0(1710)$~\cite{PDG02} later, we parameterize the ratio 
of the asymptotic matrix elements of $A_{K}$ to $A_{\pi}$ as 
$Z={\langle{K^+|A_{K^+}|S^*}\rangle 
/\langle{\pi^+|A_{\pi^+}|S^*}\rangle}$.   

We now look for the values of the above parameters which reproduce 
the measured decay branching ratios for hadronic two body decays of 
charm mesons. To this, we take 
$f_+^{(\bar KD)}(0)=0.74\pm 0.03$~\cite{PDG96} as before. 
All the other parameters mentioned before are treated as adjustable 
ones with restrictions that $a_1$ and $a_2$ should not be very far 
from the ones estimated by the perturbative QCD~\cite{perturbation}, 
the non-resonant strong phases between $-90^\circ$ and $90^\circ$, 
four-quark meson masses not very far from the ones estimated in 
Ref.~\cite{Jaffe}. When we take the following values of the 
parameters, we can reproduce the measured branching ratios for 
two body decays of charm mesons as seen in Table~4: 
the coefficients, $a_1=0.825$ and $a_2=-0.159$ ({\it c.f.}, 
$a_1^{\rm BSW}\simeq 1.09$ and $a_2^{\rm BSW}\simeq -0.09$ 
at $\mu\simeq m_c$); 
the asymptotic matrix elements of $\tilde H_w$, 
$\langle{\pi^+|\tilde H_w|D_s^+}\rangle=0.0641\times 10^{-5}$ 
(GeV)$^2$;  
the parameters describing the pole contributions of the heavier class 
of four-quark $[qq][\bar q\bar q]$ and $(qq)(\bar q\bar q)$ mesons, 
the glue-rich scalar and the hypothetical hybrid mesons, 
$k_a^*=0.0771$, $k_s^*=-0.0217$, $f_g=0.0387$, $k_H=-0.014$, 
respectively, where we have assigned the $f_0(1710)$ with the mass 
$m_{f_0}=1710$ MeV and the width $\Gamma_{f_0}=125$ MeV~\cite{PDG02} 
to the glue-rich scalar meson and have taken the ratio, 
$Z\simeq 1.56$, from the measured decay branching ratios of 
$f_0(1710)$;  
the relative phase between the factorized and non-factorizable 
amplitudes, $\delta=-19.9^\circ$;  
the phases arising from non-resonant meson-meson interactions,  
$\delta_0(\pi\pi)=\delta_0(K\bar K)=57.0^\circ$, 
$\delta_1(K\bar K)=58.7^\circ$, 
$\delta_{1/2}(\pi K)=84.4^\circ,\,\,\delta_{3/2}(\pi K)=-26.7^\circ$; 
the masses and widths of scalar non-$\{q\bar q\}$ mesons, 
$m_{\hat\sigma^*}=1.514$ GeV, $m_{E_{\pi\pi}^*}=2.164$ GeV, 
$m_{\sigma_H}=2.012$ GeV (the iso-singlet scalar hybrid) with the mass 
differences, $\Delta_s=0.1$ GeV and $\Delta_c=1.3$ GeV, and 
$\Gamma_{[qq][\bar q\bar q]}=0.198$ GeV, 
$\Gamma_{(qq)(\bar q\bar q)}=0.256$ GeV, 
$\Gamma_{\{q\bar qg\}}=0.0456$ GeV. 
The above mass values of four-quark mesons are a little higher than 
the ones in Ref.~\cite{Jaffe} while the masses of hybrid mesons with 
a normal $J^{P(C)}=0^{+(+)}$ have been much higher than the 
ones predicted by the covariant oscillator quark model~\cite{Ishida}. 
To solve the puzzle, Eq.(\ref{eq:puzzle}), the role of the glue-rich 
scalar $f_0(1710)$ and the four-quark $\hat\sigma^{s*}$ have been very 
important. 

In summary we have studied charmed scalar mesons. Since the mass of 
the newly observed $D_{s0}^+(2.32)$ has been much lower than the ones 
calculated by using various models, many different assignments (in 
addition to the ordinary scalar $\{c\bar s\}$ meson or the chiral 
partner of $D_s^+$ prior to the observation) of it have been 
proposed. These models have extra scalar mesons in addition to the 
ordinary scalar $\{c\bar q\}$ mesons. Among these models, we have 
assigned the $D_{s0}^+(2.32)$ to the $I_3=0$ member, $\hat{F}_I^+$, 
of the iso-triplet $\hat{F}_I^+$'s which belong to the lighter class 
of four-quark $[cq][\bar q\bar q]$ mesons and have investigated the 
decay rates of the members of the same multiplet. 
As the consequence, we have predicted that the iso-triplet 
$\hat{F}_I$'s are narrow and the iso-doublet, $\hat D$, mesons are 
a little broader. However, another iso-doublet $\hat D^s$'s are 
around the threshold of the decay, $\hat D^s\rightarrow D\eta$, 
which has been expected to be the dominant decay of $\hat D^s$, 
so that its rate would be very small even if it is allowed. 
The iso-singlet $\hat{F}_0^+$ is extremely narrow since its main 
decay would proceed through iso-spin violating interactions. 
The exotic state $\hat E^0$ would decay through weak interactions 
if its mass is close to $m_{\hat{F}_I}$. 

In addition to the $[cq][\bar q\bar q]$, we have the ordinary scalar 
mesons, $D_0^{*}$ and $D_{s0}^{*+}$. Their masses have been expected 
to be around $\sim 2.35$ GeV and $\sim 2.45$ GeV, respectively. 
Comparing with the $K_0^{*}$ which has been considered to be the 
$^3P_0\,\,\{n\bar s\}$, we have obtained 
$\Gamma_{D_0^{*}}\sim (60 - 90)$ MeV and 
$\Gamma_{D_{s0}^{*}}\sim (40 - 70)$ MeV. 
Therefore, we expect that there coexist two scalar mesons, i.e., the 
four-quark $\hat D$ and the ordinary $D_0^{*}$ in the region of the 
broad bump around 2.31 GeV in the $D\pi$ mass distribution which has 
been observed by the BELLE collaboration, and hope that the BELLE 
collaboration will reanalyze their data on the $D\pi$ mass 
distribution by using a model amplitude with at least two scalar 
meson poles. The strange counterpart, $D_{s0}^{*+}$, of the $D_0^{*}$ 
is massive enough to decay into the $DK$ final state so that it has 
been expected to be observed in the $DK$ channel. It is also awaited 
that experiments with high luminosities and high resolutions will 
search for a scalar resonance with a mass $\sim (2.4 - 2.5)$ GeV and 
a width $\sim (60 - 90)$ MeV in the $DK$ channel.  

Four-quark mesons have been classified into various flavor multiplets 
with different $J^{P(C)}$ and color configurations. However, it is 
expected, from the results in Ref.~\cite{Jaffe}, that the multiplets 
other than the low lying $[cq][\bar q\bar q]$ with 
$J^{P(C)}=0^{+(+)}$ and dominantly ${\bf \bar 3}\times{\bf 3}$ of 
color $SU_c(3)$ considered above are much heavier than the low lying 
ones and the $P$-wave $\{c\bar q\}$'s, i.e., $D_0^{*}$, $D_1^{*}$, 
$D_1^{'*}$, $D_2^{*}$ and $D_{s0}^{*}$, $D_{s1}^{*}$, $D_{s1}^{'*}$,
$D_{s2}^{*}$. Therefore they do not disturb the known $P$-wave 
$\{c\bar q\}$ spectrum. Namely, only the low lying scalar 
$[cq][\bar q\bar q]$ mesons coexist with the $^3P_0\,\,\{c\bar q\}$ 
mesons in the the $\sim (2.2 - 2.5)$ GeV region. 

Finally, it has been discussed that four-quark mesons can play a 
very important role in hadronic weak decays of charm mesons, in 
particular, can solve the long standing puzzle, Eq.(\ref{eq:puzzle}), 
in a overall consistent way. Therefore, hadronic weak interactions of 
charm mesons are intimately related to hadron spectroscopy and 
confirmation of the existence of four-quark mesons will open a new 
window of hadron physics, not only hadron spectroscopy but also 
hadronic weak interactions. 
\section*{Acknowledgments}
The author would like to thank Professor T.~Kunihiro and 
Professor T.~Onogi, for discussions and encouragements.  
This work is supported in part by the Grant-in-Aid for 
Science Research, Ministry of Education, Science and Culture, Japan 
(No. 13135101). 


\end{document}